\pdfoutput=1

\documentclass[12pt,a4paper]{article}

\usepackage{ifthen} 
\newboolean{pdflatex}
\setboolean{pdflatex}{true} 

\newboolean{articletitles}
\setboolean{articletitles}{true} 

\newboolean{uprightparticles}
\setboolean{uprightparticles}{false} 

\newboolean{inbibliography}
\setboolean{inbibliography}{false} 

\usepackage[top=1in, bottom=1.25in, left=1in, right=1in]{geometry}

\usepackage{microtype}
\usepackage{lineno}  
\usepackage{xspace} 
\usepackage{caption} 

\usepackage{graphicx}  
\usepackage{color}
\usepackage{colortbl}
\graphicspath{{./figs/}} 

\usepackage{amsmath} 
\usepackage{amssymb}
\usepackage{amsfonts}
\usepackage{upgreek} 

\newcommand*\patchAmsMathEnvironmentForLineno[1]{%
\expandafter\let\csname old#1\expandafter\endcsname\csname #1\endcsname
\expandafter\let\csname oldend#1\expandafter\endcsname\csname
end#1\endcsname
 \renewenvironment{#1}%
   {\linenomath\csname old#1\endcsname}%
   {\csname oldend#1\endcsname\endlinenomath}%
}
\newcommand*\patchBothAmsMathEnvironmentsForLineno[1]{%
  \patchAmsMathEnvironmentForLineno{#1}%
  \patchAmsMathEnvironmentForLineno{#1*}%
}
\AtBeginDocument{%
\patchBothAmsMathEnvironmentsForLineno{equation}%
\patchBothAmsMathEnvironmentsForLineno{align}%
\patchBothAmsMathEnvironmentsForLineno{flalign}%
\patchBothAmsMathEnvironmentsForLineno{alignat}%
\patchBothAmsMathEnvironmentsForLineno{gather}%
\patchBothAmsMathEnvironmentsForLineno{multline}%
\patchBothAmsMathEnvironmentsForLineno{eqnarray}%
}

\usepackage{hyperref}    
\usepackage[all]{hypcap} 

\usepackage{xspace} 
\usepackage{upgreek}

\def\lhcb {\mbox{LHCb}\xspace}

\def\MagUp {\mbox{\em Mag\kern -0.05em Up}\xspace}

\ifthenelse{\boolean{uprightparticles}}%
{

 \def\Pmu         {\ensuremath{\upmu}\xspace}

 \def\Ppi         {\ensuremath{\uppi}\xspace}                 
                  
 \def\Prho        {\ensuremath{\uprho}\xspace}

 \def\Ppsi        {\ensuremath{\uppsi}\xspace}

 \def\PDelta      {\ensuremath{\Delta}\xspace}                 
 \def\PXi      {\ensuremath{\Xi}\xspace}                 
 \def\PLambda      {\ensuremath{\Lambda}\xspace}                 
 \def\PSigma      {\ensuremath{\Sigma}\xspace}                 
 \def\POmega      {\ensuremath{\Omega}\xspace}                 
 \def\PUpsilon      {\ensuremath{\Upsilon}\xspace}                 
 
 \def\PB      {\ensuremath{\mathrm{B}}\xspace}                 
                  
 \def\PD      {\ensuremath{\mathrm{D}}\xspace}

 \def\PJ      {\ensuremath{\mathrm{J}}\xspace}                 
 \def\PK      {\ensuremath{\mathrm{K}}\xspace}

 \def\Pb      {\ensuremath{\mathrm{b}}\xspace}                 
 \def\Pc      {\ensuremath{\mathrm{c}}\xspace}

 \def\Pi      {\ensuremath{\mathrm{i}}\xspace}

}
{

 \def\Pmu         {\ensuremath{\mu}\xspace}

 \def\Ppi         {\ensuremath{\pi}\xspace}                 
                  
 \def\Prho        {\ensuremath{\rho}\xspace}

 \def\Ppsi        {\ensuremath{\psi}\xspace}                 
                  
 \mathchardef\PDelta="7101
 \mathchardef\PXi="7104
 \mathchardef\PLambda="7103
 \mathchardef\PSigma="7106
 \mathchardef\POmega="710A
 \mathchardef\PUpsilon="7107
                  
 \def\PB      {\ensuremath{B}\xspace}                 
                  
 \def\PD      {\ensuremath{D}\xspace}

 \def\PJ      {\ensuremath{J}\xspace}                 
 \def\PK      {\ensuremath{K}\xspace}

 \def\Pb      {\ensuremath{b}\xspace}                 
 \def\Pc      {\ensuremath{c}\xspace}

 \def\Pi      {\ensuremath{i}\xspace}

}

\makeatletter
\ifcase \@ptsize \relax
  \newcommand{\miniscule}{\@setfontsize\miniscule{4}{5}}
\or
  \newcommand{\miniscule}{\@setfontsize\miniscule{5}{6}}
\or
  \newcommand{\miniscule}{\@setfontsize\miniscule{5}{6}}
\fi
\makeatother

\DeclareRobustCommand{\optbar}[1]{\shortstack{{\miniscule (\rule[.5ex]{1.25em}{.18mm})}
  \\ [-.7ex] $#1$}}

\def\mup        {{\ensuremath{\Pmu^+}}\xspace}
\def\mun        {{\ensuremath{\Pmu^-}}\xspace} 

\def\cquark    {{\ensuremath{\Pc}}\xspace}

\def\bquark    {{\ensuremath{\Pb}}\xspace}

\def\pion   {{\ensuremath{\Ppi}}\xspace}

\def\pip    {{\ensuremath{\pion^+}}\xspace}
\def\pim    {{\ensuremath{\pion^-}}\xspace}
\def\pipm   {{\ensuremath{\pion^\pm}}\xspace}
\def\pimp   {{\ensuremath{\pion^\mp}}\xspace}

\def\rhomeson {{\ensuremath{\Prho}}\xspace}

\def\kaon    {{\ensuremath{\PK}}\xspace}
  \def\Kbar    {{\kern 0.2em\overline{\kern -0.2em \PK}{}}\xspace}

\def\KorKbar    {\kern 0.18em\optbar{\kern -0.18em K}{}\xspace}
\def\Kz      {{\ensuremath{\kaon^0}}\xspace}
\def\Kzb     {{\ensuremath{\Kbar{}^0}}\xspace}
\def\Kp      {{\ensuremath{\kaon^+}}\xspace}
\def\Km      {{\ensuremath{\kaon^-}}\xspace}
\def\Kpm     {{\ensuremath{\kaon^\pm}}\xspace}

\def\KS      {{\ensuremath{\kaon^0_{\mathrm{ \scriptscriptstyle S}}}}\xspace}

  \def\Dbar    {{\kern 0.2em\overline{\kern -0.2em \PD}{}}\xspace}
\def\D       {{\ensuremath{\PD}}\xspace}

\def\DorDbar    {\kern 0.18em\optbar{\kern -0.18em D}{}\xspace}
\def\Dz      {{\ensuremath{\D^0}}\xspace}
\def\Dzb     {{\ensuremath{\Dbar{}^0}}\xspace}

\def\Dm      {{\ensuremath{\D^-}}\xspace}
\def\Dpm     {{\ensuremath{\D^\pm}}\xspace}

\def\Dstar   {{\ensuremath{\D^*}}\xspace}

\def\Dstarp  {{\ensuremath{\D^{*+}}}\xspace}

\def\B       {{\ensuremath{\PB}}\xspace}
\def\Bbar    {{\ensuremath{\kern 0.18em\overline{\kern -0.18em \PB}{}}}\xspace}

\def\BorBbar    {\kern 0.18em\optbar{\kern -0.18em B}{}\xspace}

\def\Bu      {{\ensuremath{\B^+}}\xspace}
\def\Bub     {{\ensuremath{\B^-}}\xspace}
\def\Bp      {{\ensuremath{\Bu}}\xspace}
\def\Bm      {{\ensuremath{\Bub}}\xspace}
\def\Bpm     {{\ensuremath{\B^\pm}}\xspace}

\def\jpsi     {{\ensuremath{{\PJ\mskip -3mu/\mskip -2mu\Ppsi\mskip 2mu}}}\xspace}

  \def\Y#1S{\ensuremath{\PUpsilon{(#1S)}}\xspace}

\def\Lbar        {{\ensuremath{\kern 0.1em\overline{\kern -0.1em\PLambda}}}\xspace}
\def\LorLbar    {\kern 0.18em\optbar{\kern -0.18em \PLambda}{}\xspace}

\def\to                 {\ensuremath{\rightarrow}\xspace}

\def\CP                {{\ensuremath{C\!P}}\xspace}

\def\AT#1     {\ensuremath{A_{\mathrm{T}}^{#1}}\xspace}           

\def\C#1      {\ensuremath{\mathcal{C}_{#1}}\xspace}                       
\def\Cp#1     {\ensuremath{\mathcal{C}_{#1}^{'}}\xspace}                    
\def\Ceff#1   {\ensuremath{\mathcal{C}_{#1}^{\mathrm{(eff)}}}\xspace}        
\def\Cpeff#1  {\ensuremath{\mathcal{C}_{#1}^{'\mathrm{(eff)}}}\xspace}       
\def\Ope#1    {\ensuremath{\mathcal{O}_{#1}}\xspace}                       
\def\Opep#1   {\ensuremath{\mathcal{O}_{#1}^{'}}\xspace}                    

\newcommand{\tev}{\ifthenelse{\boolean{inbibliography}}{\ensuremath{~T\kern -0.05em eV}\xspace}{\ensuremath{\mathrm{\,Te\kern -0.1em V}}}\xspace}
\newcommand{\gev}{\ensuremath{\mathrm{\,Ge\kern -0.1em V}}\xspace}
\newcommand{\mev}{\ensuremath{\mathrm{\,Me\kern -0.1em V}}\xspace}
\newcommand{\kev}{\ensuremath{\mathrm{\,ke\kern -0.1em V}}\xspace}
\newcommand{\ev}{\ensuremath{\mathrm{\,e\kern -0.1em V}}\xspace}
\newcommand{\gevc}{\ensuremath{{\mathrm{\,Ge\kern -0.1em V\!/}c}}\xspace}
\newcommand{\mevc}{\ensuremath{{\mathrm{\,Me\kern -0.1em V\!/}c}}\xspace}
\newcommand{\gevcc}{\ensuremath{{\mathrm{\,Ge\kern -0.1em V\!/}c^2}}\xspace}
\newcommand{\gevgevcccc}{\ensuremath{{\mathrm{\,Ge\kern -0.1em V^2\!/}c^4}}\xspace}
\newcommand{\mevcc}{\ensuremath{{\mathrm{\,Me\kern -0.1em V\!/}c^2}}\xspace}

\def\mum  {\ensuremath{{\,\upmu\mathrm{m}}}\xspace}

\def\invfb   {\ensuremath{\mbox{\,fb}^{-1}}\xspace}

\def\gsim{{~\raise.15em\hbox{$>$}\kern-.85em
          \lower.35em\hbox{$\sim$}~}\xspace}
\def\lsim{{~\raise.15em\hbox{$<$}\kern-.85em
          \lower.35em\hbox{$\sim$}~}\xspace}

\def\pt         {\mbox{$p_{\mathrm{ T}}$}\xspace}

\def\pythia     {\mbox{\textsc{Pythia}}\xspace}

\def\tell1  {TELL1\xspace}
\def\ukl1   {UKL1\xspace}

\usepackage{cite} 
\usepackage{mciteplus}

\usepackage{longtable} 

\def\Aprod{\ensuremath{\mathcal{A}_{\rm prod}}\xspace}
\def\Ap{\ensuremath{\mathcal{A}_{\rm prod}}\xspace}
\def\ApBu{\ensuremath{\mathcal{A}_{\rm prod}(\Bp)}\xspace}
\def\Adet{\ensuremath{\mathcal{A}_{\rm det}}\xspace}

\def\AdetKpi{\ensuremath{\mathcal{A}_{\rm det}^{K\pi}}\xspace}
\def\Adetpi{\ensuremath{\mathcal{A}_{\rm det}^{\pi}}\xspace}
\def\AdetTIS{\ensuremath{\mathcal{A}_{\rm det}^{\rm TIS}}\xspace}
\def\AdetPID{\ensuremath{\mathcal{A}_{\rm det}^{\rm PID}}\xspace}
\def\deltaAdetKpi{\ensuremath{\delta \mathcal{A}_{\rm det}^{K\pi}}\xspace}
\def\Araw{\ensuremath{\mathcal{A}_{\rm raw}}\xspace}

\def\ArawJpsiK{\ensuremath{\mathcal{A}_{\rm raw}^{\psi K}}\xspace}
\def\ArawDpi{\ensuremath{\mathcal{A}_{\rm raw}^{D\pi}}\xspace}
\def\Acp{\ensuremath{\mathcal{A}_{\CP}}\xspace}
\def\AcpDpi{\ensuremath{\mathcal{A}_{\CP}^{D\pi}}\xspace}
\def\AcpJpsiK{\ensuremath{\mathcal{A}_{\CP}(\Bp \to \jpsi \Kp)}\xspace}

\def\ACPRESULT{\ensuremath{\left(0.09 \pm 0.27 \pm 0.07\right) \times 10^{-2}}\xspace}

\begin{document}

\renewcommand{\thefootnote}{\fnsymbol{footnote}}
\setcounter{footnote}{1}


\begin{titlepage}
\pagenumbering{roman}

\vspace*{-1.5cm}
\centerline{\large EUROPEAN ORGANIZATION FOR NUCLEAR RESEARCH (CERN)}
\vspace*{1.5cm}
\noindent
\begin{tabular*}{\linewidth}{lc@{\extracolsep{\fill}}r@{\extracolsep{0pt}}}
\ifthenelse{\boolean{pdflatex}}
{\vspace*{-2.7cm}\mbox{\!\!\!\includegraphics[width=.14\textwidth]{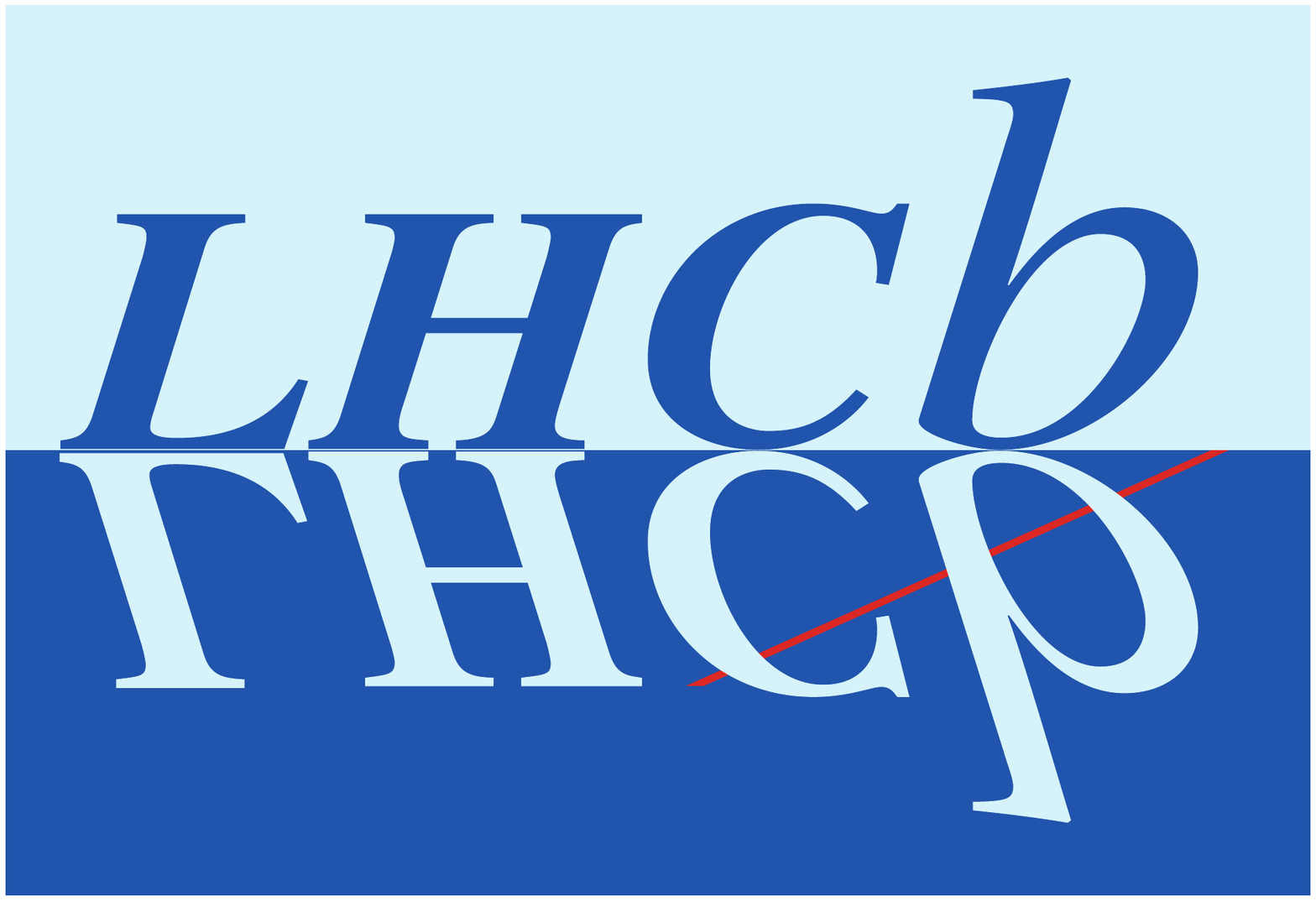}} & &}%
{\vspace*{-1.2cm}\mbox{\!\!\!\includegraphics[width=.12\textwidth]{lhcb-logo.eps}} & &}%
\\
 & & CERN-EP-2016-325 \\  
 & & LHCb-PAPER-2016-054 \\  
 & & 19 January 2017 \\ 
 & & \\
\end{tabular*}

\vspace*{1.0cm}

{\normalfont\bfseries\boldmath\huge
\begin{center}
Measurement of the $\Bpm$ production asymmetry
and the \CP asymmetry in $\Bpm \to \jpsi \Kpm$ decays
\end{center}
}

\vspace*{1.0cm}

\begin{center}
The LHCb collaboration\footnote{Authors are listed at the end of this paper.}
\end{center}

\vspace{\fill}

\begin{abstract}
  \noindent
   The \Bpm meson production asymmetry in $pp$ collisions is measured using $\Bp \to \Dzb \pip$ 
   decays.
   The data were recorded by the LHCb experiment 
   during Run~1 of the LHC at centre-of-mass energies of $\sqrt{s}=$ 7 and 8\tev.
   The production asymmetries, integrated over transverse momenta in the range 
   $2 < \pt < 30$\gevc, and rapidities in the range 
   $2.1 < y < 4.5$, are measured to be
   \begin{align*}
   \Ap(\Bp,\sqrt{s}=7\tev) &= (-0.41 \pm 0.49\pm 0.10)\times 10^{-2},\\
   \Ap(\Bp,\sqrt{s}=8\tev) &= (-0.53 \pm 0.31\pm 0.10)\times 10^{-2},
  \end{align*}
   where the first uncertainties are statistical and the second are systematic.
  These production asymmetries are used to correct the raw asymmetries of $\Bp \to \jpsi \Kp$ decays,
  thus allowing a measurement of the \CP asymmetry,
  \begin{equation*}
  \AcpJpsiK = \ACPRESULT.
  \end{equation*}
  
\end{abstract}

\vspace*{2.0cm}

\begin{center}
  Submitted to Phys.~Rev.~D
\end{center}

\vspace{\fill}

{\footnotesize 
\centerline{\copyright~CERN on behalf of the \lhcb collaboration, licence \href{http://creativecommons.org/licenses/by/4.0/}{CC-BY-4.0}.}}
\vspace*{2mm}

\end{titlepage}


\newpage
\setcounter{page}{2}
\mbox{~}
%
%
%
%



\renewcommand{\thefootnote}{\arabic{footnote}}
\setcounter{footnote}{0}



\pagestyle{plain} 
\setcounter{page}{1}
\pagenumbering{arabic}


%


\section{Introduction}
\label{sec:Introduction}

One of the primary goals of the LHCb experiment is to search for effects of physics beyond the Standard Model 
through measurements of \CP-violating asymmetries in beauty- and charm-hadron decays.
A challenge for such measurements in $pp$ collisions is that
the heavy flavour production rates differ between particles
and antiparticles.
These production asymmetries cannot be precisely predicted 
since they arise in the non perturbative \bquark or \cquark quark hadronisation process~\cite{Norrbin:2000zc,Norrbin:2000jy,Norrbin:1999by}.
The effects of production asymmetries cancel in measurements of the difference between \CP asymmetries
of two different decays of the same hadron species.

\begin{figure}[h]
  \begin{center}
    \includegraphics*[width=0.48\textwidth]{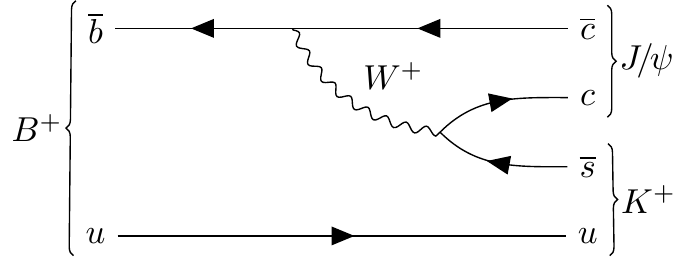}\hfill 
    \includegraphics*[width=0.48\textwidth]{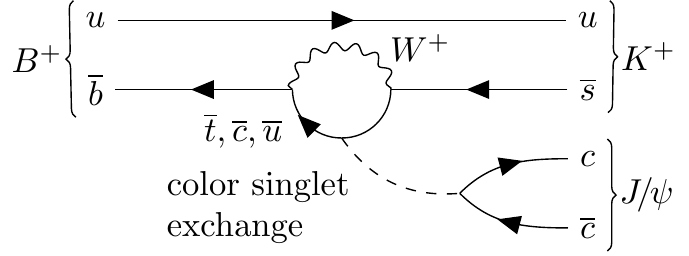}    
  \caption{Tree and loop (penguin) diagrams for the $\Bp \to \jpsi \Kp$ decay.}
  \label{fig:feynman}
  \end{center}
\end{figure}

The \CP asymmetries of \Bu meson decay rates\footnote{The inclusion of charge-conjugate processes is implied throughout, except in the discussion of asymmetries.} are often measured relative to that of the decay $\Bp \to \jpsi \Kp$.
The leading tree-level diagram for this decay, shown in Fig.~\ref{fig:feynman} (left), is colour-suppressed
and the total decay amplitude may receive a sizeable contribution from the gluonic loop diagram shown in Fig.~\ref{fig:feynman} (right).
Therefore, the $\Bp \to \jpsi \Kp$ decay can in principle exhibit a \CP asymmetry due to the interference between these amplitudes.
The current world average value of the \CP asymmetry is $\AcpJpsiK = (0.3 \pm 0.6)\%$~\cite{patrignani2016review}
and the uncertainty represents a limitation in many $\Bu$ meson $\CP$ asymmetry measurements that use this channel
as a reference.

This analysis exploits the decay $\Bp \to \Dzb \pip$, which is dominated by 
a Cabibbo- and colour-favoured tree-level amplitude and is therefore expected to have a \CP asymmetry
with a smaller value and uncertainty than for the $\Bp \to \jpsi \Kp$ mode.
The $\Bp \to \Dzb \pip$ decay mode is used to measure the production asymmetry between the cross-sections for 
$\Bm$ and $\Bp$ mesons, defined as
\begin{equation}
\label{Eq:BuAp}
\Ap(\Bp) \equiv \frac{\sigma(\Bm) - \sigma(\Bp)}{\sigma(\Bm) + \sigma(\Bp)}.
\end{equation}
Since the production asymmetry is expected to be a function of the kinematics,
the measurement is performed
in nine bins of \Bp transverse momentum, \pt, and rapidity, $y$, 
within the fiducial region $2 < \pt < 30$\gevc and $2.1 < y < 4.5$.
Measurements are performed on two data sets corresponding to 
integrated luminosities of 1\invfb and 2\invfb, 
recorded at centre-of-mass energies of 7 and 8~TeV in 2011 and 2012, respectively.
These measurements complement the existing LHCb studies of heavy flavour production
asymmetries~\cite{LHCb-PAPER-2012-009,LHCb-PAPER-2012-026,Aaij:2014bba,LHCb-PAPER-2014-053}.
A combined analysis of $\Bp \to \Dzb \pip$ and $\Bp \to \jpsi \Kp$ decays
allows a measurement of the \CP asymmetry in the latter mode.
The raw charge asymmetry for a flavour-specific decay to the final state
$f$ ($\bar{f}$) accessible in decays of $B^-$ ($B^+$) mesons is defined as
\begin{equation}
\begin{aligned}
\label{Eq:Araw}
\Araw(\Bp\to\bar{f}) &\equiv \frac{N(\Bm \to f) - N(\Bp \to \bar{f})}{N(\Bm \to f) + N(\Bp \to \bar{f})}.
\end{aligned}
\end{equation}
For the two decay modes under study, the asymmetries are well approximated by
\begin{equation}
\begin{aligned}
\label{Eq:BuAp}
\Araw(\Bp \to \Dzb \pip) &= \Ap(\Bp) + \Adet(\Dzb \pip) + \Acp(\Bp \to \Dzb \pip),\\
\Araw(\Bp \to \jpsi \Kp) &= \Ap(\Bp) + \Adet(\jpsi \Kp) + \Acp(\Bp \to \jpsi \Kp),
\end{aligned}
\end{equation}
where $\Adet$ is the detector-induced asymmetry resulting from differences in 
the detection efficiencies between particles and antiparticles.
All contributions to $\Adet$ are measured on independent control samples from the same data set.
The high correlation of $\Adet$ between the two decay modes 
implies a partial cancellation in their difference.
This cancellation and the low level of \CP violation in the $\Bp \to \Dzb \pip$ decay mode
enable a precise measurement of $\AcpJpsiK$.

\section{The LHCb detector}
\label{sec:Detector}

The \lhcb detector~\cite{Alves:2008zz,LHCb-DP-2014-002} is a single-arm forward
spectrometer covering the pseudorapidity range $2<\eta <5$,
designed for the study of particles containing \bquark or \cquark
quarks. The detector includes a high-precision tracking system
consisting of a silicon-strip vertex detector surrounding the $pp$
interaction region, a large-area silicon-strip detector located
upstream of a dipole magnet with a bending power of about 
$4{\mathrm{\,Tm}}$, and three stations of silicon-strip detectors and straw
drift tubes placed downstream of the magnet.
Data samples corresponding to roughly equal integrated luminosities were recorded
with configurations in which the magnetic field was pointing vertically upwards and downwards.
This largely cancels any charge asymmetries in the reconstruction efficiency for 
charged particles.
The tracking system provides a measurement of momentum, $p$, of charged particles with
a relative uncertainty that varies from 0.5\% at low momentum to 1.0\% at 200\gevc.
The minimum distance of a track to a primary vertex (PV), the impact parameter (IP),
is measured with a resolution of $(15+29/\pt)\mum$,
where \pt is the component of the momentum transverse to the beam, in\,\gevc.
Different types of charged hadrons are distinguished using information
from two ring-imaging Cherenkov detectors.
Muons are identified by a
system composed of alternating layers of iron and multiwire
proportional chambers.
The online event selection is performed by a trigger~\cite{LHCb-DP-2012-004},
which consists of a hardware stage, based on information from the calorimeter and muon
systems, followed by a two-stage software trigger, which applies a full event
reconstruction.
This analysis makes use of inclusive dimuon and beauty selections at the software trigger stages.


\section{Selection of \boldmath{$\Bp \to \Dzb \pip$} decays}
\label{sec:SelectionDpi}

The selection of signal candidate $\Bp \to \Dzb \pip$ decays closely follows a recent LHCb
analysis involving the same decay channel~\cite{LHCb-PAPER-2016-003}.
Events are considered for the analysis if they contain 
a track with large enough \pt and IP
to satisfy the requirements of the first stage of the software trigger.
An inclusive beauty selection is applied at the second stage of the software trigger.
Candidate $\Dzb \to \Kp\pim(\Dzb \to \Kp\pip\pim\pim)$ decays are constructed
from the intersection of two (four) tracks that satisfy appropriate kaon or pion 
particle identification (PID) criteria,
and that have a large \pt and significant IP
with respect to all primary vertices.
These candidates must have a mass within $\pm 25$\mevcc of the $\Dzb$
mass~\cite{patrignani2016review}.
Each \Dzb candidate is combined with a high \pt track that is identified as a pion to create a displaced vertex that is consistent with a decay of a $\Bp$ meson.
The $\Bp$ candidates are required to have a mass within the range $5079$--$5899$\mevcc.
To reduce to a negligible level the uncertainty related to L0 trigger asymmetries,
 it is explicitly required that a positive L0 trigger decision was
caused by a particle that is distinct from any of the final-state particles
that compose the signal candidate. 
This requirement is independent of whether or not the signal candidate itself
also caused a positive L0 trigger decision and is therefore 
referred to as triggering independently of signal (TIS)~\cite{LHCb-DP-2012-004}.

For both the two- and four-body \Dzb-mode selections, a pair of 
boosted decision tree (BDT) discriminators~\cite{Breiman}, 
implementing the gradient boost algorithm~\cite{Roe}, is used to achieve further background suppression.
The first of these BDTs is trained to reject candidates with fake \Dzb decays,
and the second to reject backgrounds with real \Dzb decays.
The BDTs are trained using simulated $\Bp \to \Dzb \pip$ signal decays 
and a sample of decays from data with 
masses in the range \mbox{$5900$--$7200$\mevcc} to model the combinatorial background in the nominal mass range.
For the training of the first BDT, a background sample is provided by candidates
with \Dzb masses that differ by more than $\pm 30$\mevcc from the known \Dzb mass.
The second BDT is trained using a background sample of candidates with \Dzb masses within $\pm25$\mevcc of the known \Dzb mass.
A loose cut on the classifier response of the first BDT is applied prior to training the second one. 
The inputs to the BDTs include properties of each particle ($p$, \pt, and the IP significance) 
and additional properties of the \B and \Dz composite particles 
(decay time, flight distance, decay vertex quality, radial distance between 
the decay vertex and the PV, and the angle between the reconstructed momentum vector 
and the line connecting the production and decay vertex). 
A further input to the BDTs is an isolation variable
\begin{equation}
I_{\pt} = \frac{\pt(\Bpm) - \Sigma \pt}{\pt(\Bpm) + \Sigma \pt},
\end{equation}
for which the sum is taken over tracks that are not part of the signal candidate
but fall within a cone of half-angle $\Delta R < 1.5$~radians, where $(\Delta R)^2 = (\Delta\theta)^2 + (\Delta\phi)^2$,
and $\Delta\theta$ and $\Delta\phi$ are the differences in polar and azimuthal angle of each track with respect to the \Bp candidate direction. Tracks are only considered in the isolation cone 
if they are associated, by smallest IP, to the same primary vertex as the signal candidate.
Signal decays are expected to have larger values of $I_{\pt}$ than background.

The cut on the second BDT response is optimised by minimising the expected uncertainty 
on the asymmetry between the yields of $\Bm \to \Dz \pim$ and $\Bp \to \Dzb \pip$. 
No PID information is used in the BDT training, but the purity of the sample is further improved by requiring all kaon and pion candidates to satisfy PID criteria. 
Events  containing more than one $\Bp \to \Dzb \pip$ candidate amount to less than 1 \%, and in these cases
the candidate with the highest quality \Bp decay vertex is selected.

The raw asymmetries between the yields of $\Bm \to \Dz \pim$ and $\Bp \to \Dzb \pip$ 
decays are determined by binned maximum likelihood fits to the mass distributions of 
selected \Bm and \Bp candidates, treating the two- and four-body \Dzb modes separately.
The fit function is built from a signal component and three background components.
A sum of two Gaussian functions with asymmetric power-law tails and an additional Gaussian function 
are combined to model $\Bp \to \Dzb \pip$ decays~\cite{LHCb-PAPER-2016-003}. 
Misidentified $\Bp \to \Dzb \Kp$ decays
have a distribution that is below the signal peak 
with a tail that extends to lower masses.
They are modelled by the sum of two Gaussian functions with asymmetric power-law tail components.
Partially reconstructed decays with an additional particle from a \Dstar or \rhomeson meson decay
form a background at masses lower than that of the signal peak. 
This component is described by a combination of analytical functions with shapes that
depend on the spin-parity of the missing particle, following the method described in Ref.~\cite{LHCb-PAPER-2016-003}. 
A linear function is adequate to describe the combinatorial background distribution. 
The yield of misidentified $\Bp \to \Dzb \Kp$ decays is constrained 
with an independent control sample of these decays, combined with the calibrated particle identification
efficiencies and misidentification rates~\cite{Anderlini:2202412}.
With the exception of the tail parameters, which are fixed to values obtained from simulation, 
all parameters are allowed to vary in the fit.

\begin{figure}[ht]
  \begin{center}
    \includegraphics*[width=0.95\textwidth]{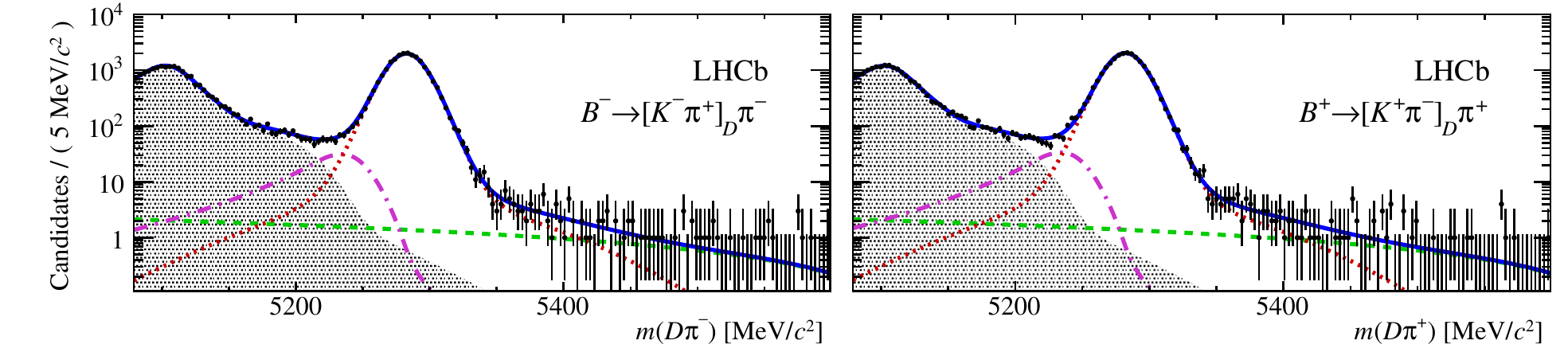}\\
     \includegraphics*[width=0.95\textwidth]{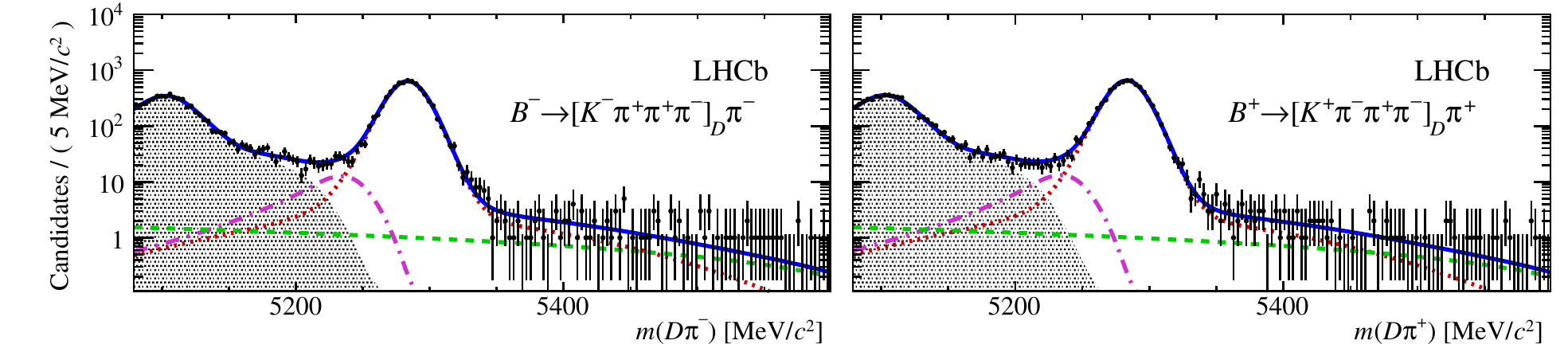}
 \caption{Mass distributions of selected (top) $\Bpm \to [\Kpm \pimp]_{D}\pi^{\pm}$
and (bottom) \mbox{$\Bpm \to [\Kpm \pipm \pimp \pimp]_{D}\pi^{\pm}$} 
candidates in the bin with $4.5 < \pt < 9.5$\gevc and $2.10 < y < 2.85$.
These distributions sum over the two centre-of-mass energies.
$B^{-}$ candidates are displayed on the left, and $B^{+}$ candidates on the right. 
The red dotted lines indicate the contribution from $\Bpm \to D\pi^{\pm}$ decays. 
The purple dash-dotted lines indicate the contribution from misidentified $\Bpm \to DK^{\pm}$ decays. 
The grey shaded regions at low values of reconstructed mass indicate the contribution from various partially reconstructed $B$ decays, and the green dashed lines indicate the combinatorial background. The total fit function is shown by the blue solid lines.
The fit in other kinematic bins is similar, aside from the specific signal and background component yields. 
 \label{fig:fit_Dpi}}
  \end{center}
\end{figure}

Figure~\ref{fig:fit_Dpi} shows the fits to the mass distributions
in the bin with $4.5 < \pt < 9.5$\gevc and $2.10 < y < 2.85$.
The subsequent analysis is based on separate fits for the nine kinematic bins
and two centre-of-mass energies.
The signal yields for each of the nine kinematic bins are listed in Table~\ref{tab:Yields}.
The \pt and $y$ intervals of each bin are defined in the second and third columns.
The yields sum over \Bpm meson charges and centre-of-mass energies.
Integrated over the fiducial acceptance, 
$2 < \pt < 30$\gevc and $2.1 < y < 4.5$,
the fits return signal yields of around $2.3\times 10^{5}$ decays for the $\Dzb \to \Kp\pim$ mode
and around $1.3\times 10^{5}$ decays for the $\Dzb \to \Kp\pip\pim\pim$ mode.

\section{Selection of \boldmath{$\Bp \to \jpsi \Kp$} decays}
\label{sec:SelectionPsiK}

The selection of $\Bp \to \jpsi \Kp$  decays with $\jpsi \to \mup\mun$ is based 
on events in which a muon or a generic track, 
with large \pt and IP, satisfies the requirements of the first-stage  software trigger. 
Events must be selected based on a dimuon signature by the second-level software trigger.
Candidate $\jpsi \to \mup\mun$ decays are reconstructed from high-\pt muon candidates
with large IP with respect to all PVs.
A mass interval of $3057$--$3127$~\mevcc is imposed on the \jpsi candidates.
These candidates are combined with a high-\pt identified kaon with a significant IP
with respect to all PVs, where the $J/\psi$ candidate invariant mass is constrained to its known value in the combination. 
The L0 trigger TIS requirement is applied in the same way as for the $\Dzb\pip$ selection. 
A single BDT classifier is used to improve the purity of the $\Bp \to \jpsi \Kp$ sample. 
This classifier is trained 
on a similar set of variables as that for the $\Bp \to \Dzb \pip$ selection, 
and exhibits very similar performance in terms of signal efficiency and background rejection. 
Events  containing more than one $\Bp \to \jpsi \Kp$ candidate amount to less than 1 \%, and in these cases
 the candidate with the highest quality \Bp decay vertex is selected.

A simultaneous fit of the mass distributions across the kinematic bins is performed, where the same value of \AcpJpsiK is assumed for all bins. 
The signal peak is described using a Gaussian function with an additional asymmetric power-law tail component.
The mean of the Gaussian is constrained to be the same in all kinematic bins,
while its width and the tail parameters are allowed to vary between bins.
A small background from misidentified $\Bu \to \jpsi \pip$ decays is described by a similar function, with fixed shape parameters taken from simulation.
The yield of this contribution is allowed to vary in each kinematic bin, 
but a single raw asymmetry is shared between all bins.
The contribution from random particle combinations is described
by a linear function.
The yield of this component and the slope parameter are allowed to vary in each kinematic bin.
The yield is also fitted separately for each \Bpm charge.

\begin{figure}[ht]
  \begin{center}
    \includegraphics*[width=0.95\textwidth]{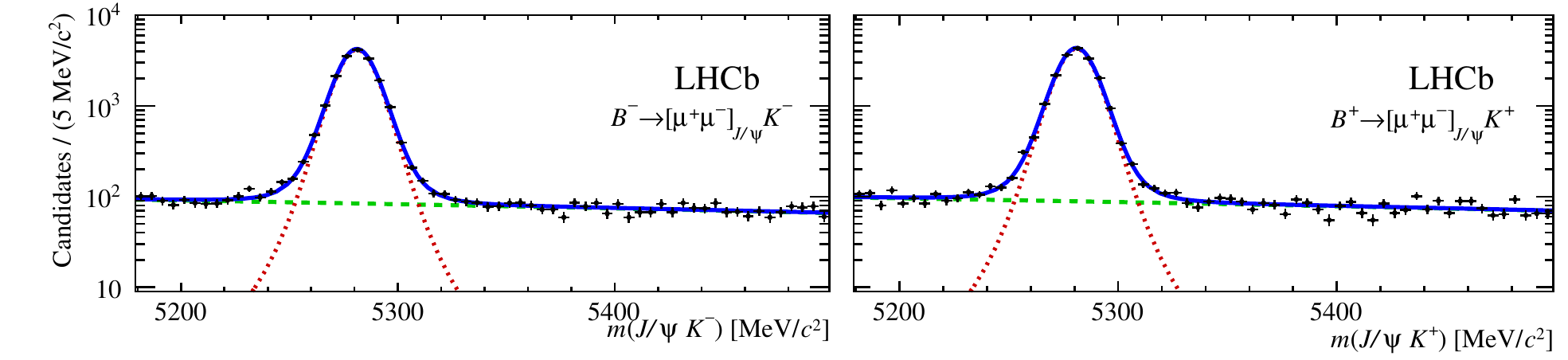}
  \caption{Mass distribution of selected $B^{\pm} \to \jpsi K^{\pm}$ candidates in the 
bin with $4.5 < \pt < 9.5$~\gevc and $2.10 < y < 2.85$.
These distributions sum over the two centre-of-mass energies.
$B^{-}$ candidates are displayed on the left, and $B^{+}$ candidates on the right. 
The signal components are displayed as red dotted lines while the background 
from combinatorial events is shown by the green dashed lines. 
The fit in other kinematic bins is similar, aside from the specific signal and background component yields. 
 \label{fig:fit_JpsiK}}
  \end{center}
\end{figure}

Integrated over the full fiducial acceptance, a signal yield of about $2.3\times 10^{5}$ events is measured. 
Table~\ref{tab:Yields} lists the yields of each signal decay mode in each of the kinematic bins
summing over the two centre-of-mass energies.
An example of the fit in the bin with $4.5 < \pt < 9.5$\gevc and $2.10 < y < 2.85$ is displayed in Fig.~\ref{fig:fit_JpsiK}. 

\begin{table}[!b!]
\centering
\caption{\label{tab:Yields}The \pt and $y$ intervals for each kinematic bin, 
  and the corresponding signal yields in each of the $B^+$ decay modes,
  summing over the two centre-of-mass energies.}
\begin{tabular}{l  r@{--}l r@{--}l  r@{~$\pm$~}l r@{~$\pm$~}l  r@{~$\pm$~}l r@{~$\pm$~}l }
\hline \noalign{\vskip 2pt}
Bin
& \multicolumn{2}{c}{\pt}
& \multicolumn{2}{c}{$y$}
& \multicolumn{4}{c}{$B^+ \to \Dzb \pi^+$} & \multicolumn{2}{c}{$B^+ \to \jpsi K^+$}  \\ 

& \multicolumn{2}{c}{[\gevc]}& \multicolumn{2}{c}{}
&\multicolumn{2}{l}{$\Dzb\to\Kp\pim$} 
&\multicolumn{2}{l}{$\Dzb\to\Kp\pim\pip\pim$} 
&\multicolumn{2}{l}{$\jpsi \to \mu^+\mu^-$}  \\[2pt]
\hline \noalign{\vskip 2pt}
1   & 2.0 & 4.5  & 2.10 & 2.85 & 13604&118 & 1549&42 & 17319&194   \\
2   & 2.0 & 4.5  & 2.85 & 3.3  & 18587&145 & 4022&66 & 26038&229   \\
3   & 2.0 & 4.5  & 3.3  & 4.5  & 19946&151 & 6347&87 & 31110&260   \\
4   & 4.5 & 9.5  & 2.10 & 2.85 & 44470&219 & 14209&131 & 34939&231 \\
5   & 4.5 & 9.5  & 2.85 & 3.3  & 47597&240 & 23895&163 & 36682&230 \\
6   & 4.5 & 9.5  & 3.3  & 4.5  & 31137&200 & 24014&170 & 31345&212 \\
7   & 9.5 & 30   & 2.10 & 2.85 & 33516&195 & 23378&167 & 25174&189 \\
8   & 9.5 & 30   & 2.85 & 3.3  & 20176&159 & 20332&151 & 15110&136 \\
9   & 9.5 & 30   & 3.3  & 4.5  & 4767&73   & 8832&97   & 8602&191  \\
\hline \noalign{\vskip 2pt}
\multicolumn{5}{c}{Integrated} & 233390&537 & 126350&393 & 226319&632 \\
\hline \noalign{\vskip 2pt}
\end{tabular}
\end{table}


\section{Measurement of the \boldmath{\Bu} production asymmetry}
\label{sec:ApBp}

\begin{table}[!t!]
\caption{\label{Tab:auto_Aprod_2011}
A summary of the terms that enter the production asymmetry determination (Eq.~\ref{Eq:BuAp}) in the 7~TeV data set.
The \pt and $y$ intervals of each bin are provided in Tab.~\ref{tab:Yields}.
The L0 trigger asymmetry \AdetTIS is omitted from this table since it is assumed to be 
independent of the \Bp kinematics.
All uncertainties are statistical.}
\centering
\resizebox{\textwidth}{!}{
\begin{tabular}{lr@{~$\pm$~}lr@{~$\pm$~}lr@{~$\pm$~}lr@{~$\pm$~}lr@{~$\pm$~}l}
\hline\noalign{\vskip 4pt}
Bin& \multicolumn{2}{c}{\ArawDpi ($\times 10^{-2}$)}& \multicolumn{2}{c}{\AcpDpi ($\times 10^{-2}$)}& \multicolumn{2}{c}{\AdetKpi ($\times 10^{-2}$)}& \multicolumn{2}{c}{\Adetpi ($\times 10^{-2}$)}& \multicolumn{2}{c}{\AdetPID ($\times 10^{-2}$)}\\
\noalign{\vskip 4pt}\hline\noalign{\vskip 4pt}
1& $-1.1$ & $1.5$& $+0.08$ & $0.05$& $-1.39$ & $0.22$& $-0.04$ & $0.13$& $-0.066$ & $0.006$\\
2& $-1.5$ & $1.3$& $+0.08$ & $0.05$& $-1.18$ & $0.15$& $-0.05$ & $0.08$& $+0.017$ & $0.017$\\
3& $-1.7$ & $1.1$& $+0.07$ & $0.05$& $-1.19$ & $0.16$& $-0.04$ & $0.09$& $+0.077$ & $0.007$\\
4& $-1.1$ & $0.8$& $+0.07$ & $0.05$& $-1.23$ & $0.21$& $+0.03$ & $0.11$& $-0.0875$ & $0.0021$\\
5& $-1.6$ & $0.7$& $+0.07$ & $0.04$& $-1.03$ & $0.13$& $+0.03$ & $0.08$& $-0.049$ & $0.004$\\
6& $-1.5$ & $0.8$& $+0.06$ & $0.04$& $-1.10$ & $0.13$& $-0.02$ & $0.08$& $+0.2092$ & $0.0033$\\
7& $-0.7$ & $0.8$& $+0.06$ & $0.04$& $-0.84$ & $0.20$& $+0.04$ & $0.13$& $-0.0606$ & $0.0026$\\
8& $-2.6$ & $0.9$& $+0.05$ & $0.04$& $-0.65$ & $0.12$& $+0.05$ & $0.12$& $+0.0645$ & $0.0022$\\
9& $-0.2$ & $1.6$& $+0.04$ & $0.04$& $-1.07$ & $0.11$& $+0.06$ & $0.12$& $+0.3951$ & $0.0032$\\
\noalign{\vskip 4pt}\hline\noalign{\vskip 4pt}
\end{tabular}

}
\end{table}

\begin{table}[!b!]
\caption{\label{Tab:auto_Aprod_2012}
A summary of the terms that enter the production asymmetry determination (Eq.~\ref{Eq:BuAp}) in the 8~TeV data set.
The L0 trigger asymmetry \AdetTIS is omitted from this table since it is assumed to be 
independent of the \Bp kinematics.
All uncertainties are statistical.}
\centering
\resizebox{\textwidth}{!}{
\begin{tabular}{lr@{~$\pm$~}lr@{~$\pm$~}lr@{~$\pm$~}lr@{~$\pm$~}lr@{~$\pm$~}l}
\hline\noalign{\vskip 4pt}
Bin& \multicolumn{2}{c}{\ArawDpi ($\times 10^{-2}$)}& \multicolumn{2}{c}{\AcpDpi ($\times 10^{-2}$)}& \multicolumn{2}{c}{\AdetKpi ($\times 10^{-2}$)}& \multicolumn{2}{c}{\Adetpi ($\times 10^{-2}$)}& \multicolumn{2}{c}{\AdetPID ($\times 10^{-2}$)}\\
\noalign{\vskip 4pt}\hline\noalign{\vskip 4pt}
1& $-0.7$ & $1.0$& $+0.08$ & $0.05$& $-1.16$ & $0.13$& $-0.17$ & $0.09$& $+0.059$ & $0.004$\\
2& $-1.2$ & $0.9$& $+0.07$ & $0.05$& $-1.08$ & $0.09$& $-0.10$ & $0.06$& $+0.0855$ & $0.0029$\\
3& $-2.8$ & $0.8$& $+0.07$ & $0.05$& $-0.93$ & $0.10$& $-0.07$ & $0.06$& $+0.0659$ & $0.0026$\\
4& $-1.3$ & $0.5$& $+0.07$ & $0.05$& $-1.07$ & $0.12$& $-0.10$ & $0.07$& $-0.0144$ & $0.0008$\\
5& $-1.7$ & $0.4$& $+0.07$ & $0.04$& $-0.99$ & $0.08$& $-0.11$ & $0.05$& $+0.0963$ & $0.0013$\\
6& $-1.2$ & $0.5$& $+0.06$ & $0.04$& $-0.79$ & $0.08$& $-0.06$ & $0.06$& $+0.1323$ & $0.0024$\\
7& $-1.0$ & $0.5$& $+0.06$ & $0.04$& $-0.93$ & $0.11$& $-0.02$ & $0.08$& $+0.0120$ & $0.0012$\\
8& $-1.0$ & $0.6$& $+0.05$ & $0.04$& $-0.78$ & $0.07$& $-0.14$ & $0.08$& $+0.0581$ & $0.0029$\\
9& $-1.8$ & $1.0$& $+0.04$ & $0.04$& $-0.56$ & $0.07$& $-0.10$ & $0.08$& $+0.0914$ & $0.0017$\\
\noalign{\vskip 4pt}\hline\noalign{\vskip 4pt}
\end{tabular}

}
\end{table}

The $\Bp$ production asymmetry is determined in the nine bins of \pt and $y$ according to
\begin{equation}
\label{Eq:BuAp}
\Ap(\Bp) = \ArawDpi - \AcpDpi - \AdetKpi  - \Adetpi - \AdetPID - \AdetTIS,
\end{equation}
where \ArawDpi and \AcpDpi are the raw charge asymmetry and \CP asymmetry in the $\Bp \to \Dzb \pip$ decay,
respectively.
The four \Adet terms correct for detector-induced asymmetries and
will be described in the following.
All terms other than \AcpDpi are
evaluated separately for the four disjoint data sets corresponding
to the two centre-of-mass energies and the two magnet polarities.
An average of the \ArawDpi values for the two $\Dzb$ decay modes is computed with 
weights that are chosen to minimise the uncertainty.
The same weights are used to compute averages over the two $\Dzb$ decay modes for all other terms in Eq.~\ref{Eq:BuAp}
apart from \AdetTIS, which is independent of the $\Dzb$ decay.
Tables~\ref{Tab:auto_Aprod_2011} and~\ref{Tab:auto_Aprod_2012} list 
the values of the first five terms in Eq.~\ref{Eq:BuAp} for the 7 and 8\tev data sets, respectively.
The overall detection asymmetry has two main contributions.
The first arises because $\Km$ mesons have a larger nuclear interaction cross-section than $\Kp$
mesons.
This means that more $\Km$ mesons than $\Kp$ mesons interact inelastically with the detector material before they 
leave enough hits to be reconstructed in the tracking stations.
The resulting $\Km$--$\Kp$ detection asymmetry is around $10^{-2}$.
The second cause of asymmetry is the different trajectories of positively and negatively
charged particles, which therefore have different sensitivities to misalignments and inhomogeneities of the detector.
This source contributes to all detection asymmetry terms.
It is partially cancelled when averaging measurements over data recorded
with the dipole magnet in the two polarities.

The \Dzb detection asymmetry, \AdetKpi, is
measured using 
samples of $\Dm$ mesons that are produced in the primary $pp$ interactions and decay to the $\Kp\pim\pim$ and $\KS\pim$ final states.
The $\KS$ mesons are reconstructed in their decay to $\pi^+\pi^-$.
Within a small phase-space region in terms of the \Dm decay products,
 it is assumed that
the detection asymmetry for a $K^+\pi^-$ pair can be determined using
\begin{equation}
\label{Eq:AKpi}
\AdetKpi = \Araw(\Dm \to \Kp\pim\pim) - \Araw(\Dm \to \KS\pim),
\end{equation}
with a small correction for the effects of \CP violation in $\Kz-\Kzb$ mixing
and the different material interactions of $\Kz$ and $\Kzb$.
For each of the $\Dm\to\Kp\pim\pim$ candidates one of the two $\pim$ mesons
is randomly labelled as being {\em matched} to the $\Bp \to \Dzb\pip$ signal.
A weight is assigned to each $\Dm\to\Kp\pim\pim$ candidate
such that the kinematic distributions of the $\Kp$ and the matched $\pim$ 
agree with those from the signal \Dzb decays.
For the  $\Dzb \to \Kp\pip\pim\pim$ sample, the procedure is repeated for each of the two possible pions with opposite charge to the kaon, averaging over the two.
Each $\Dm\to\KS\pim$ candidate is assigned a weight, 
such that the $\pim$ kinematic distributions agree with those of the {\em unmatched} $\pim$ in the 
weighted $\Dm\to\Kp\pim\pim$ sample, and the \Dm kinematic distributions are equalised between the two \Dm decay modes.
This ensures cancellation of the \Dm production asymmetry, and means that any detection asymmetry associated with the unmatched $\pim$ is cancelled with a corresponding asymmetry affecting the $\Dm\to\KS\pim$ sample.
This weighting procedure is performed for each of the nine $\Bp$ kinematic bins.
The raw asymmetries that enter Eq.~\ref{Eq:AKpi} are
determined by fitting the weighted mass spectra for the four 
combinations of $\Dpm$ decay modes and charges.

Using a detailed description of the LHCb detector and cross-section measurements from fixed target experiments~\cite{patrignani2016review}
the nuclear interaction contribution to the pion asymmetry is estimated to be negligibly small.
The tracking asymmetry can therefore be assumed to be the same for pions and muons.
The \pip tracking asymmetry, \Adetpi, is therefore inferred from that of muons measured using a sample of $\jpsi \to \mup\mun$ decays
in which one of the muons is reconstructed without requiring hits in all tracking stations~\cite{LHCb-DP-2013-002}.
Weights are assigned to the $\jpsi$ candidates such that the 
kinematic distributions of this muon match those of the $\pim$ in the $\Bp \to \Dzb\pip$ sample.

The PID requirements on the $\Bp \to \Dzb \pip$ decays
can introduce asymmetries.
Corrections are determined using a control sample of $\Dstarp \to \Dz \pip$ decays, with $\Dz \to \Km\pip$, in which 
no PID requirements are imposed on the \Km or \pip from the \Dz decay.
The asymmetry associated with PID requirements on the \Dzb decays
is partially accounted for in the \AdetKpi correction, since PID
requirements are imposed on the final state kaons and pions in the \Dm control samples.
The requirements are tighter in these control samples, 
and so a residual correction must still be applied.
The sum of this correction, and a corresponding correction for the PID requirement
on the $\pip$ from the $\Dstarp \to \Dz \pip$ decays, is denoted \AdetPID.

The asymmetry associated with the TIS trigger efficiency, \AdetTIS, 
is determined using a sample of $b$-hadron decays to the final state $\Dzb \mu^+\nu_{\mu} X$ 
with $\Dzb \to \Kp\pim$.
An unbiased probe of the TIS trigger efficiency is provided by the subset of 
these in which the muon prompted a positive decision by the L0 muon trigger.
The corresponding asymmetries do not exhibit any kinematic dependence, 
and so a single correction is determined for each centre-of-mass energy,
and is applied to all kinematic bins.
The measured \AdetTIS values are $(+0.16 \pm 0.16)\times 10^{-2}$ and $(+0.02 \pm 0.10)\times 10^{-2}$ 
for the 7~TeV and 8~TeV data sets, respectively.

The \CP asymmetry, \AcpDpi, is estimated from 
measurements of the CKM angle $\gamma$ and the hadronic parameters
of $\Bp \to \Dzb\pip$ decays~\cite{Aaij:2016kjh}.
Different values are obtained for the 
$\Dzb \to \Kp\pim$ and $\Dzb \to \Kp\pim\pip\pim$
decay modes due to the smaller coherence factor 
from the competing hadronic resonances in the four-body mode.
The asymmetries are
  \begin{align*}
\AcpDpi(K^-\pi^+) &= \left(0.09^{+0.05}_{-0.04}\right) \times 10^{-2},\\
\AcpDpi(K^-\pi^+\pi^-\pi^+) &= \left(0.00^{+0.05}_{-0.02}\right) \times 10^{-2},
  \end{align*}
with a 55\% correlation between the uncertainties on these two quantities.
The \AcpDpi values reported in Tables~\ref{Tab:auto_Aprod_2011} and~\ref{Tab:auto_Aprod_2012}
are averaged over the two- and four-body modes.
These values vary between the kinematic bins due to the different weights
of the two- and four-body modes.

Several sources of systematic uncertainty arise in the determination of the 
production asymmetries.
Their contributions are listed in Table~\ref{Tab:ApSystematics}.
Variations in the weighting procedures that are used to determine \AdetKpi
and \Adetpi yield uncertainties of $0.07 \times 10^{-2}$ and $0.04 \times 10^{-2}$, respectively.
An uncertainty of $0.04 \times 10^{-2}$ is assigned to a possible pion nuclear interaction asymmetry
that is not accounted for in the tracking efficiency measurements with muons from $\jpsi$ decays.
Finally, the \AcpDpi uncertainties are included in the total systematic uncertainty,
which is taken to be correlated between the kinematic bins.

\begin{table}[!t!]\centering
\caption{\label{Tab:ApSystematics}Systematic uncertainties on the \ApBu measurement.
The \AcpDpi uncertainty varies between the kinematic bins 
and the range is indicated.
All systematic uncertainties are considered to be correlated between
kinematic bins.}
\begin{tabular}{lc}
\noalign{\vskip 2pt}\hline\noalign{\vskip 4pt}
Source & Size ($\times 10^{-2}$) \\ 
\noalign{\vskip 2pt}\hline\noalign{\vskip 4pt}
\AdetKpi method & $\pm 0.07$ \\[2pt] 
\Adetpi method & $\pm 0.04$ \\[2pt] 
Pion nuclear interactions & $\pm 0.04$ \\[2pt] 
\AcpDpi & $\pm (0.04-0.05)$ \\[2pt] 
\noalign{\vskip 2pt}\hline\noalign{\vskip 2pt}
\end{tabular}
\end{table}

\begin{table}[!b!]\centering
\caption{\label{Tab:auto_Aprod}The measured \Ap values for each kinematic bin
and integrated over the full kinematic acceptance, $2 < \pt < 30$\gevc and $2.1 < y < 4.5$.
The integrated values sum over the asymmetries in each bin,
weighted by the values, $w$, in the second and fourth columns for the two centre-of-mass energies.
The first uncertainty is the statistical uncertainty on \ArawDpi and is uncorrelated between
the kinematic bins.
The second uncertainty is the statistical uncertainty on the detection asymmetry corrections 
and is taken to be correlated between the kinematic bins.
The third uncertainty is purely systematic and is assumed to be correlated between bins.}
\resizebox{\textwidth}{!}{
\begin{tabular}{l c c  c c }
\noalign{\vskip 4pt}\hline\noalign{\vskip 4pt}
Bin & \multicolumn{1}{c}{$w(7\tev)$} & $\Aprod(\Bp, 7\tev)$ ($\times 10^{-2}$)& \multicolumn{1}{c}{$w(8\tev)$} & $\Aprod(\Bp, 8\tev)$ ($\times 10^{-2}$) \\
\noalign{\vskip 4pt}\hline\noalign{\vskip 4pt}
1& 0.182& $+0.12 \pm 1.54 \pm 0.30 \pm 0.10$& 0.174& $+0.42 \pm 0.96 \pm 0.19 \pm 0.10$\\
2& 0.092& $-0.54 \pm 1.25 \pm 0.24 \pm 0.10$& 0.088& $-0.15 \pm 0.89 \pm 0.14 \pm 0.10$\\
3& 0.156& $-0.78 \pm 1.13 \pm 0.24 \pm 0.10$& 0.156& $-1.95 \pm 0.75 \pm 0.16 \pm 0.10$\\
4& 0.208& $-0.04 \pm 0.78 \pm 0.29 \pm 0.10$& 0.202& $-0.22 \pm 0.50 \pm 0.17 \pm 0.10$\\
5& 0.094& $-0.78 \pm 0.70 \pm 0.22 \pm 0.10$& 0.095& $-0.83 \pm 0.45 \pm 0.14 \pm 0.10$\\
6& 0.144& $-0.82 \pm 0.80 \pm 0.22 \pm 0.10$& 0.151& $-0.61 \pm 0.52 \pm 0.14 \pm 0.10$\\
7& 0.064& $-0.04 \pm 0.79 \pm 0.28 \pm 0.10$& 0.068& $-0.17 \pm 0.51 \pm 0.17 \pm 0.10$\\
8& 0.028& $-2.24 \pm 0.92 \pm 0.23 \pm 0.10$& 0.030& $-0.19 \pm 0.60 \pm 0.15 \pm 0.10$\\
9& 0.032& $+0.23 \pm 1.59 \pm 0.23 \pm 0.10$& 0.038& $-1.33 \pm 1.05 \pm 0.14 \pm 0.10$\\
\noalign{\vskip 4pt}\hline\noalign{\vskip 4pt}
\multicolumn{2}{l}{Integrated}& $-0.41 \pm 0.42 \pm 0.26 \pm 0.10$& \multicolumn{1}{l}{}& $-0.53 \pm 0.26 \pm 0.16 \pm 0.10$\\
\noalign{\vskip 4pt}\hline\noalign{\vskip 4pt}
\end{tabular}

}
\end{table}

\begin{figure}[!t!]
\includegraphics[width=0.49\textwidth]{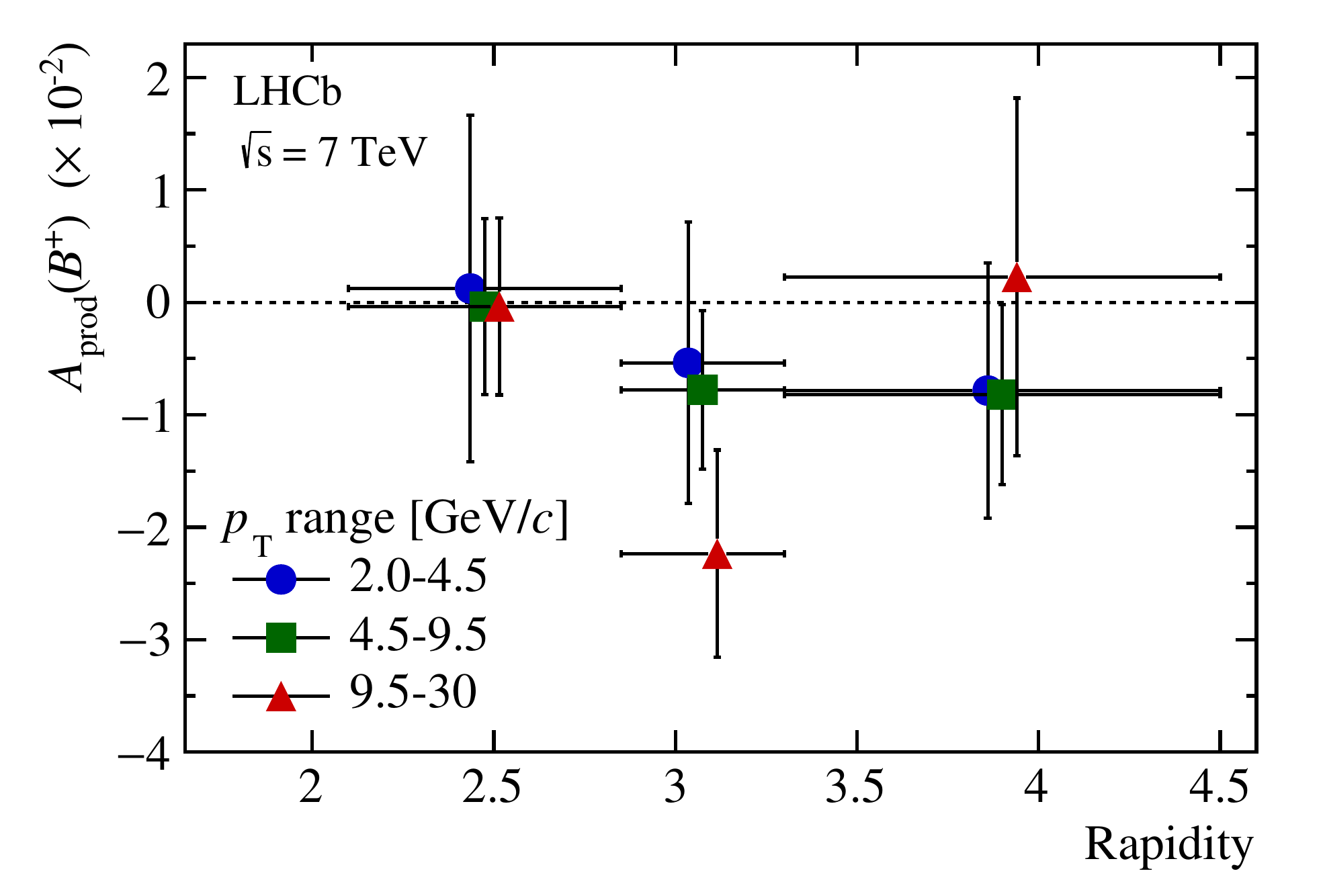}
\includegraphics[width=0.49\textwidth]{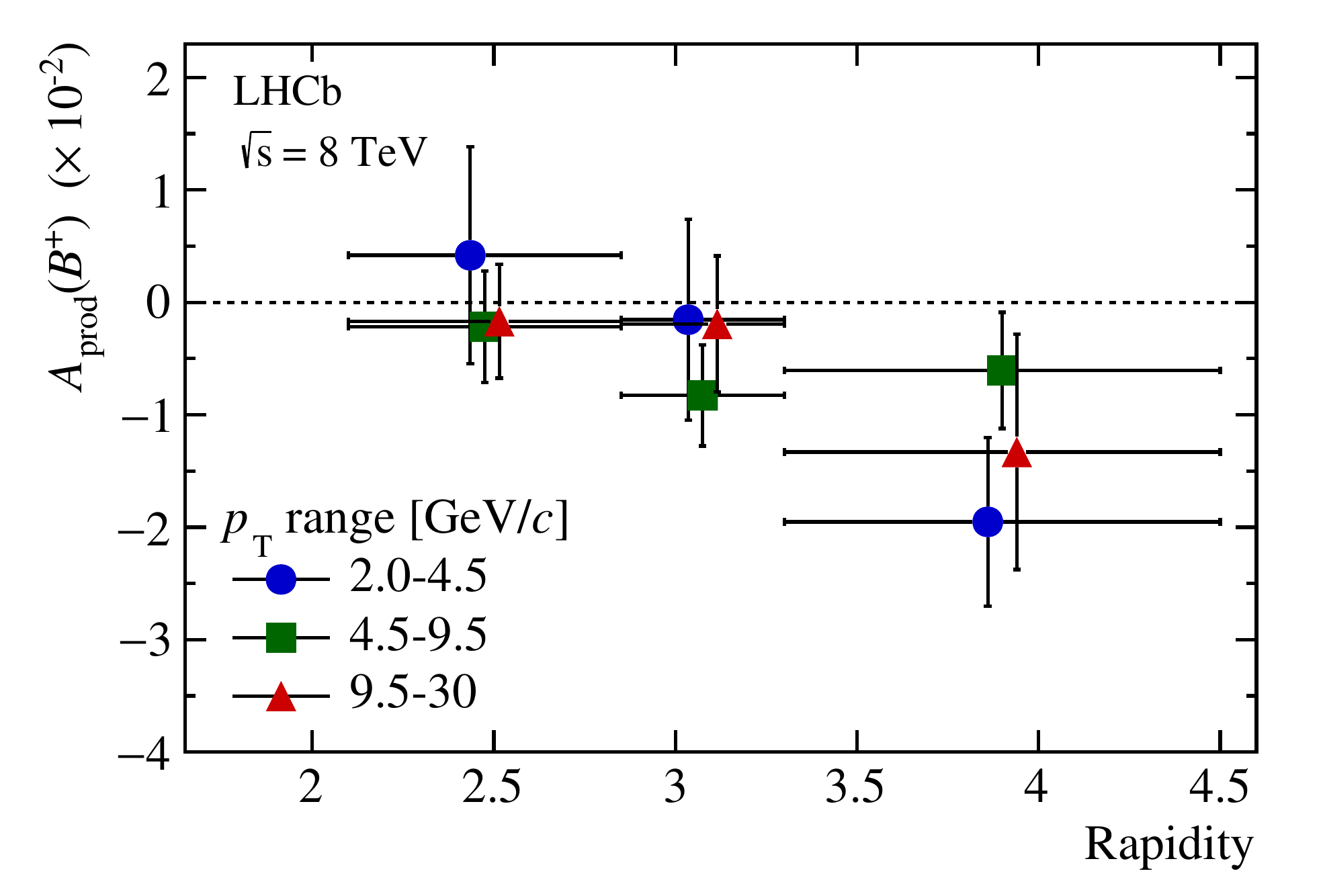}
\caption{\label{Fig:ApResult}The measured \ApBu as a function
of rapidity of the $B$ meson in three bins of $\pt$.
The ranges of \pt are indicated in the legends.
The left- and right-hand figures correspond to 7 and 8~TeV centre-of-mass energies, respectively.}
\end{figure}

The measured $\Ap(\Bp)$ values for each kinematic bin are listed in 
Table~\ref{Tab:auto_Aprod} for both centre-of-mass energies.
They are shown as a function of rapidity for the three \pt ranges in Fig.~\ref{Fig:ApResult}.
Samples of simulated \Bpm decays are produced using \pythia~8~\cite{Sjostrand:2007gs,Sjostrand:2006za} with a specific LHCb configuration~\cite{LHCb-PROC-2010-056},
and are used to determine the weights that are assigned to each of the nine bins, 
such that the sum corresponds to the asymmetry integrated over the full fiducial region covering
$2 < \pt < 30$~\gevc and $2.1 < y < 4.5$.
These weights are listed in Table~\ref{Tab:auto_Aprod}.
The integrated asymmetries, which are also reported in Table~\ref{Tab:auto_Aprod}, are
  \begin{align*}
   \Ap(\Bp,\sqrt{s}=7\tev) &= (-0.41 \pm 0.49\pm 0.10)\times 10^{-2},\\
   \Ap(\Bp,\sqrt{s}=8\tev) &= (-0.53 \pm 0.31\pm 0.10)\times 10^{-2},
  \end{align*}
where the first uncertainty is statistical and includes contributions
from \ArawDpi and the detection asymmetry corrections which are inherently statistical
in nature. The second uncertainty is systematic.
Several cross-checks are performed. 
The measured value of $\Ap(\Bp)$ is found to have no statistically significant
dependence on the \Bp decay time or kaon momentum.
Statistically compatible results are obtained for the two magnet polarities.

\section{Measurement of \boldmath{\AcpJpsiK}}
\label{sec:ACP}

The value of \AcpJpsiK is determined according to
\begin{equation}
\label{Eq:AcpJpsiK}
\AcpJpsiK = \ArawJpsiK - \deltaAdetKpi -\ArawDpi + \AcpDpi,
\end{equation}
where \ArawJpsiK is the raw asymmetry of $\Bpm \to \jpsi \Kpm$ decays
and \deltaAdetKpi corrects for the different detection asymmetries
of the two decay modes.
The two final states differ by the transformation of a $\pi^+\pi^-$ pair to a $\mu^+\mu^-$ pair, where the only significant contribution to the difference between the overall detection asymmetries arises from the charged kaon asymmetry.
The method used to determine \AdetKpi, as described in the previous section, is applied to the $\jpsi K^+$ final state
 by considering the muon with opposite charge to the kaon as a pion. 
The difference between this and the corresponding asymmetry for the $\Bu \to \Dzb \pip$ mode
is defined as $\deltaAdetKpi = \AdetKpi(B \to \jpsi K) - \AdetKpi(B \to \Dzb\pi)$.
The uncertainties are cancelled to a large degree in this difference.
Table~\ref{tab:deltaAKpi} lists the values of \deltaAdetKpi for each kinematic bin.
The values of \deltaAdetKpi are positive, since the kaons in the $\jpsi K^+$ decays tend to have 
higher momenta than those in the $\Bu \to \Dzb \pip$ decays.
A further asymmetry could result from differences between the kinematic distributions of the 
pion in the $\Bu \to \Dzb \pip$ decay compared to the $\mu^+$ in the $\jpsi K^+$ decay, but this is estimated to be negligibly small. 

\begin{table}[!ht!]\centering
\caption{\label{tab:deltaAKpi}Residual differences \deltaAdetKpi, measured in each bin of $B$ kinematics. These are the effective values after
summing over centre-of-mass energies and averaging over the two \Dzb decay modes.}
\begin{tabular}{lc}
\noalign{\vskip 4pt}\hline\noalign{\vskip 4pt}
Bin& \deltaAdetKpi ($\times 10^{-2}$)\\
\noalign{\vskip 4pt}\hline\noalign{\vskip 4pt}
1& 0.15 $\pm$ 0.04\\
2& 0.22 $\pm$ 0.03\\
3& 0.24 $\pm$ 0.05\\
4& 0.26 $\pm$ 0.02\\
5& 0.29 $\pm$ 0.02\\
6& 0.21 $\pm$ 0.02\\
7& 0.27 $\pm$ 0.02\\
8& 0.23 $\pm$ 0.01\\
9& 0.05 $\pm$ 0.02\\
\noalign{\vskip 4pt}\hline\noalign{\vskip 4pt}
\end{tabular}

\end{table}

The values of \ArawJpsiK in each bin are corrected according to Eq.~\ref{Eq:AcpJpsiK} using measurements of \ArawDpi, \deltaAdetKpi and \AcpDpi in order to extract \AcpJpsiK. Gaussian constraints are applied to the values of \ArawDpi and \deltaAdetKpi, such that the statistical uncertainty on these parameters is included in the overall statistical uncertainty for \AcpJpsiK. 
A systematic uncertainty of $0.02\times 10^{-2}$ is assigned for the use of fixed parameters in the  mass fits, while a systematic uncertainty of $0.05\times 10^{-2}$ is assigned for the method used to measure \deltaAdetKpi. 
The \AcpDpi values contribute a systematic uncertainty of $0.04\times 10^{-2}$.
The final result is
\begin{align*}
\AcpJpsiK = \ACPRESULT,
\end{align*}
where the first uncertainty is statistical and the second is systematic. 
By fixing all Gaussian constrained parameters to have zero uncertainty, 
the contribution from the finite $\Bp \to \jpsi\Kp$ statistics is found to be $\pm 0.20 \times 10^{-2}$.
This result is consistent with, and improves upon, the current world average value of 
$\AcpJpsiK = (0.3 \pm 0.6)\%$~\cite{patrignani2016review}.

\section{Summary and conclusions}

   The \Bu meson production asymmetry is a crucial input in the measurement of \CP asymmetries in \Bu decays.
   A sample of $\Bp \to \Dzb \pip$ decays is used to measure the production asymmetry.
   The analysed data set corresponds to integrated luminosities of
   1 and 2\invfb recorded during 2011 and 2012 at proton-proton
   centre-of-mass energies of 7 and 8\tev, respectively. 
   The production asymmetries are measured in nine bins of transverse momenta and rapidity,
   covering the region $2 < \pt < 30$\gevc and $2.1 < y < 4.5$, and separately for the two 
   centre-of-mass energies.
   The measurements are generally consistent with zero asymmetry within typical uncertainties
   of roughly $10^{-2}$, which is in agreement with \bquark-quark hadronisation models~\cite{Norrbin:2000zc,Norrbin:2000jy,Norrbin:1999by}.
   Integrated over the full \pt and $y$ ranges, the production asymmetries are measured to be
  \begin{align*}
   \Ap(\Bp,\sqrt{s}=7~\tev) &= (-0.41 \pm 0.49\pm 0.10)\times 10^{-2},\\
   \Ap(\Bp,\sqrt{s}=8~\tev) &= (-0.53 \pm 0.31\pm 0.10)\times 10^{-2},
  \end{align*}
  where the first uncertainty accounts for all statistical sources, 
and the second accounts for all systematic sources.
  A simultaneous study of the $\Bu \to \jpsi K^{+}$ decay allows a measurement of its \CP asymmetry,
  \begin{equation*}
  \AcpJpsiK = \ACPRESULT.
  \end{equation*}

\section*{Acknowledgements}
 
\noindent We express our gratitude to our colleagues in the CERN
accelerator departments for the excellent performance of the LHC. We
thank the technical and administrative staff at the LHCb
institutes. We acknowledge support from CERN and from the national
agencies: CAPES, CNPq, FAPERJ and FINEP (Brazil); NSFC (China);
CNRS/IN2P3 (France); BMBF, DFG and MPG (Germany); INFN (Italy); 
FOM and NWO (The Netherlands); MNiSW and NCN (Poland); MEN/IFA (Romania); 
MinES and FASO (Russia); MinECo (Spain); SNSF and SER (Switzerland); 
NASU (Ukraine); STFC (United Kingdom); NSF (USA).
We acknowledge the computing resources that are provided by CERN, IN2P3 (France), KIT and DESY (Germany), INFN (Italy), SURF (The Netherlands), PIC (Spain), GridPP (United Kingdom), RRCKI and Yandex LLC (Russia), CSCS (Switzerland), IFIN-HH (Romania), CBPF (Brazil), PL-GRID (Poland) and OSC (USA). We are indebted to the communities behind the multiple open 
source software packages on which we depend.
Individual groups or members have received support from AvH Foundation (Germany),
EPLANET, Marie Sk\l{}odowska-Curie Actions and ERC (European Union), 
Conseil G\'{e}n\'{e}ral de Haute-Savoie, Labex ENIGMASS and OCEVU, 
R\'{e}gion Auvergne (France), RFBR and Yandex LLC (Russia), GVA, XuntaGal and GENCAT (Spain), Herchel Smith Fund, The Royal Society, Royal Commission for the Exhibition of 1851 and the Leverhulme Trust (United Kingdom).

\addcontentsline{toc}{section}{References}
\setboolean{inbibliography}{true}
\bibliographystyle{LHCb}
\bibliography{main,LHCb-PAPER,LHCb-CONF,LHCb-DP,LHCb-TDR}

\newpage
\centerline{\large\bf LHCb collaboration}
\begin{flushleft}
\small
R.~Aaij$^{40}$,
B.~Adeva$^{39}$,
M.~Adinolfi$^{48}$,
Z.~Ajaltouni$^{5}$,
S.~Akar$^{59}$,
J.~Albrecht$^{10}$,
F.~Alessio$^{40}$,
M.~Alexander$^{53}$,
S.~Ali$^{43}$,
G.~Alkhazov$^{31}$
% ...

{\footnotesize \it
% ...
$ ^{5}$Clermont Universit{\'e}, Universit{\'e} Blaise Pascal, CNRS/IN2P3, LPC, Clermont-Ferrand, France\\
% ...
$ ^{10}$Fakult{\"a}t Physik, Technische Universit{\"a}t Dortmund, Dortmund, Germany\\
% ...
$ ^{31}$Petersburg Nuclear Physics Institute (PNPI), Gatchina, Russia\\
% ...
$ ^{39}$Universidad de Santiago de Compostela, Santiago de Compostela, Spain\\
$ ^{40}$European Organization for Nuclear Research (CERN), Geneva, Switzerland\\
% ...
$ ^{43}$Nikhef National Institute for Subatomic Physics, Amsterdam, The Netherlands\\
% ...
$ ^{48}$H.H. Wills Physics Laboratory, University of Bristol, Bristol, United Kingdom\\
% ...
$ ^{53}$School of Physics and Astronomy, University of Glasgow, Glasgow, United Kingdom\\
% ...
$ ^{59}$University of Cincinnati, Cincinnati, OH, United States\\
% ...
}
\end{flushleft}

\end{document}